\documentclass[aip,jap,reprint,a4paper,superscriptaddress]{revtex4-1}

\usepackage{graphicx}
\usepackage{natbib}
\usepackage{amsmath}
\usepackage{amssymb}
\usepackage{color}
\usepackage{esint}
\usepackage{subfigure}
\usepackage{soul}
\usepackage{color}
\usepackage{xcolor}
\definecolor{Red}{rgb}{0.9,0.0,0.1}
\usepackage[colorlinks=true,citecolor=blue,linkcolor=blue,hyperindex]{hyperref}
\usepackage{lineno}

\begin{document}


\title{Indirect switching of vortex polarity through magnetic  dynamic coupling}



\author{G.B.M. Fior}
\affiliation{Laborat\'orio Nacional de Luz S\'{\i}ncrotron, 13083-970, Campinas, SP, Brazil}
\author{E.R.P. Novais}
\altaffiliation[Present address: ]{Instituto de F\'{\i}sica, Universidade Federal de Alagoas, 57072-970, Macei\'{o}, AL, Brazil}
\affiliation{Centro Brasileiro de Pesquisas F\'{\i}sicas, 22290-180,  Rio de Janeiro, RJ, Brazil}

\author{J.P. Sinnecker}
\affiliation{Centro Brasileiro de Pesquisas F\'{\i}sicas, 22290-180,  Rio de Janeiro, RJ, Brazil}
\author{A.P. Guimar\~aes}
\affiliation{Centro Brasileiro de Pesquisas F\'{\i}sicas, 22290-180,  Rio de Janeiro, RJ, Brazil}
\author{F. Garcia}
\affiliation{Centro Brasileiro de Pesquisas F\'{\i}sicas, 22290-180,  Rio de Janeiro, RJ, Brazil}
\date{\today}
\begin{abstract}
Magnetic vortex cores exhibit a gyrotropic motion, and may reach a critical velocity, at which point they invert their z-component of the magnetization. We performed micromagnetic simulations to describe this vortex core polarity reversal in magnetic nanodisks presenting a perpendicular anisotropy.  We found that the critical velocity decreases with increasing perpendicular anisotropy, therefore departing from a universal criterion, that relates this velocity only to the exchange stiffness of the material. This leads to a critical velocity inversely proportional to the vortex core radius. We have also shown that in a pair of interacting disks, it is possible to switch the core vortex polarity through a non-local excitation; exciting one disk by applying a rotating magnetic field, one is able to switch the polarity of a neighbor disk, with a larger perpendicular anisotropy.
\end{abstract}
\maketitle

\section{Introduction}
In recent years, a great deal of interest has been given to low-dimensional magnetic systems, in particular systems with magnetic vortices found in their equilibrium magnetic configuration.\cite{Guimaraes:2009,Shinjo:2000,Cowburn:1999,Chien:2007} The vortices are characterized by presenting curling magnetization in the plane - the curling direction defines the circulation: $c=+1$ (CCW) or $-1$ (CW). In the center of the structure - the vortex core - the magnetization points out of the plane, up or down, defining the polarity $p$ $(p=+1\ {\rm or}\ p=-1)$. The core is surrounded by a circular region of opposite magnetization (dip).
 The core radius, as well as other properties, can be tailored by varying the perpendicular magnetic anisotropy of the nanostructures.\cite{Garcia:2010,Novais:2011,Novais:2013}


The dynamic behavior of the magnetic vortices has been studied extensively. \cite{Yamada:2007,Kim:2008,Bohlens:2008,Ruotolo:2009,Garcia:2012}
When the vortex core is displaced from the equilibrium position, two relevant consequences are the asymmetry of the dip and the appearance of a restoring force. This force is responsible for the motion - gyrotropic motion - toward the disk center. This motion has an eigenfrequency, which depends on the ratio of the thickness to the radius of the disk $\beta= L/R$. For small thickness disks $\omega_0\propto M_{\rm s}\beta$,\cite{Guslienko:2006} where $M_{\rm s}$ is the saturation magnetization of the disk material. However, the influence of uniaxial perpendicular anisotropy on the dynamics of vortex cores has not been extensively investigated so far.

Among the dynamic properties studied, the switching of vortex cores polarity has deserved special attention in the literature. For example, the polarity can be inverted through several methods, which include the application of pulsed in-plane fields and polarized spin currents, among others.\cite{Waeyenberge:2006,Yamada:2007,Lee:2007,Lee:2008,Kim:2008,Loubens:2009} It has been observed that rotating magnetic fields at the resonance frequency ($\omega_0$) are particularly effective in performing this inversion.


Some proposed applications for vortex cores involve the interaction between side by side
vortices, as, for example, the vortex-based transistor. \cite{Kumar:2014}
Two disks with magnetic vortices in the ground state configuration interact very weakly, since they present magnetic flux closure. However, a vortex core performing a gyrotropic motion generates sufficient magnetostatic energy to dynamically couple to a neighbor vortex, as demonstrated in some recent studies.\cite{Mesler:2007,Bocklage:2008,Liu:2009,Vogel:2010,Sugimoto:2011,Jung:2011,Sukhostavets:2013,Garcia:2012} Particularly interesting is the fact that it is possible to transfer energy, with negligible loss between two neighbor vortices by stimulated gyrotropic motion.\cite{Vogel:2010}  This coupling is strongly dependent on the  distance $d$ between the centers of the vortices, and on their relative polarities.\cite{Vogel:2011,Garcia:2012,Sukhostavets:2013,Sinnecker:2014}

The aim of this work is to demonstrate, using micromagnetic simulation, that it is possible to switch remotely the polarity of a given vortex, through a non-local excitation of a neighbor vortex; this was reached by suitably combining the perpendicular magnetic anisotropies of the nanostructures.
In order to achieve this goal we initially investigated the influence of the uniaxial perpendicular magnetic anisotropy ($K_z$) on the dynamic properties of isolated disks, particularly the switching of the vortex core, induced by a rotating magnetic field in the plane of the disk. We showed that the eigenfrequency, the velocity (critical velocity $v_{\rm crit}$) of the core immediately before polarity reversal, as well as its distance (critical distance $r_{\rm crit}$) from the center, depend on the intensity of this $K_z$.

All micromagnetic simulations were carried out using the freely-available software OOMMF\cite{oommf}, employing permalloy disks with $2R=500$\,nm and thickness $L= 20$\,nm. The magnetic parameters are $M_s=860$\,kA/m for the saturation magnetization, the exchange stiffness is $A=13$ pJ/m, the Gilbert damping $\alpha=0.01$. The cell size is $5\times 5 \times 5$\,nm$^3$. We varied the perpendicular anisotropies from $K_{z}=0\ \rm{to}\ 237$\,kJ/m$^3$; beyond this upper limit the vortex structure is no longer stable for the disk dimension studied and a skyrmion configuration is the state of minimum energy. \cite{Garcia:2010,Novais:2011}
\begin{figure}[h]
\includegraphics[width=0.9\columnwidth]{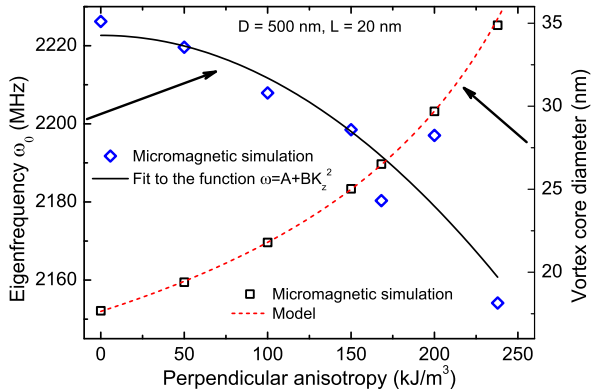}
\caption{Dependence of the vortex eigenfrequency $\omega_0$  on the perpendicular anisotropy $K_{z}$, for a disk of $2R = 500$\,nm and thickness $L = 20$\,nm (blue diamonds, left-hand scale); the line is a fit to the function $\omega_0=A+BK_z^2$. Also shown the dependence of the vortex core diameter on the perpendicular anisotropy (black squares, right-hand scale); dotted line given by model in Ref \cite{Garcia:2010}.}
\label{fig:OmegaVsKz}
\end{figure}
\section{Reversal of the vortex polarity of a single disk as a function of $K_z$}
\label{sec:ReversalPolarity}
To study the switching of a vortex core induced by another vortex driven by a resonant rotating magnetic field, we initially had to investigate the properties of individual vortices as a function of $K_z$. As we had previously demonstrated,\cite{Garcia:2010, Novais:2011,Novais:2013} the vortex core is very sensitive to the $K_z$ intensity, i.e., the larger it is, the larger will be the vortex core diameter: $R_c=\sqrt{2A/\mu_0 M_{\rm s}^2-2K_{z}}$. This dependence is represented on the right-hand axis of Fig.~\ref{fig:OmegaVsKz}. Hence, since one expects that the gyrotropic frequency would depend on the core diameter\cite{Guslienko:2006}, we also expected to find a dependence of the gyrotropic frequency $\omega_0$ on the value of the $K_z$. We could then determine the dependence of the gyrotropic frequency on $K_{z}$.



We obtained $\omega_0$ as a function of the value of the $K_z$ by observing the relaxation of an excited vortex toward the equilibrium position. The dependence of this frequency on $K_z$ is given by the left-hand axis of Fig.~\ref{fig:OmegaVsKz}; with increasing anisotropy, $\omega_0$ falls by some 3\% in the range of  $K_{z}$ studied. The fact that this range is relatively narrow is relevant for the study conducted with the pair of vortices: a study of the excitation of a vortex by a rotating magnetic field demonstrated that this field was effective in a frequency range of about 6\%.\cite{Kim:2007}
\begin{figure}[h]
\includegraphics[width=0.9\columnwidth]{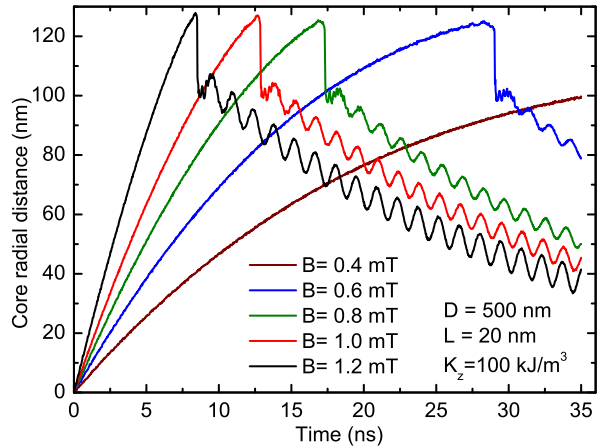}
\caption{Temporal evolution of the vortex core distance from the center of the disk, for several rotating magnetic field amplitudes, for a disk of $2R=500$\,nm, thickness $L=20$\,nm and anisotropy constant $K_{z}=100$\,kJ/m$^3$.}
\label{fig:CoreRadialDistance}
\end{figure}

When a vortex core is removed from the center equilibrium position by a rotating planar field with a frequency close to $\omega_0$, it moves away from the center in a spiral motion. The tangential core velocity will increase as the core moves away from the center. Above a certain threshold field amplitude $B_{\rm thres}$, it will reach a critical velocity value $v_{\rm crit}$ at which the polarization is reverted.\cite{Guslienko:2008}  It has been demonstrated that above this threshold, the distance $r_{\rm crit}=v_{\rm crit}/\omega_0(K_{z})$ from the center where the inversion of the vortex core occurs, is independent of the field amplitude. \cite{Yamada:2007,Kim:2008,Lee:2008} On the other hand, for amplitudes below or equal to $B_{\rm thres}$, the vortex core reaches a stationary orbit,  i.e., it does not switch polarity.

We illustrate the independence of the critical velocity on the field amplitude in Fig.~\ref{fig:CoreRadialDistance}, that shows the time evolution of the vortex core distance from the center, for several rotating field amplitudes, at frequency $\omega_0(K_{z}=100$\,kJ/m$^3$).  The sudden drops in the curves occur at $r_{\rm crit}$; after the inversion, the core motion changes direction, and begins to turn in a direction opposite to the rotating magnetic field, therefore leaving the resonance condition. The core then relaxes toward the center of the disk. We also observe that $r_{\rm crit}$ is the same for all field amplitudes, but for larger amplitudes, as expected, the vortex core reaches this radius earlier.

\begin{figure}[h]
\includegraphics[width=0.9\columnwidth]{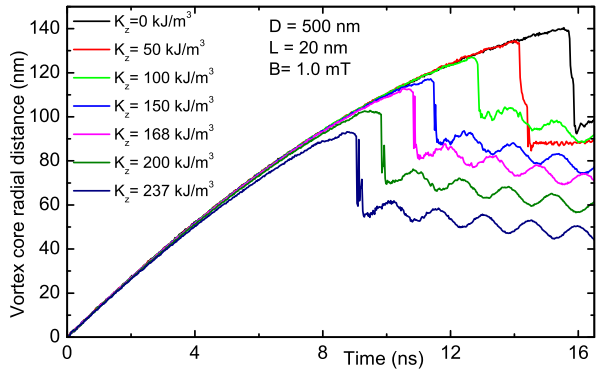}
\caption{Temporal evolution of the distance between the vortex core and the center of the disk, for field amplitude B=1 mT,  for a disk of $D = 500$\,nm and thickness $L = 20$\,nm and different values of the perpendicular anisotropy constant $K_{z}$.}
\label{fig:CoreRadialDistanceXkz}
\end{figure}

Magnetic rotating fields with the gyrotropic frequencies $\omega_0(K_{z})$ were used to excite the disk with a given $K_{z}$. For each value of $K_z$ the simulation resulted in a different critical velocity and radius, as shown in Fig.~\ref{fig:CoreRadialDistanceXkz}. It is worth noting that all the curves in Fig. \ref{fig:CoreRadialDistanceXkz} coincide below $v_{\rm crit}$, independently of the value of the anisotropy, i.e., the radial component of the core velocity is the same, independent of $K_z$.  It is clear from Fig.~\ref{fig:CoreRadialDistanceXkz} that by including a $K_z$ term, the radial distance traveled by the vortex core before reversing the polarity is smaller for larger $K_z$. Thus, a magnetic system with $K_z$ has a critical radius smaller than the system without anisotropy. Analyzing Fig.~\ref{fig:CoreRadialDistanceXkz} we can obtain the critical radius as a function of the $K_z$. The radius $r_{\rm crit}$ decreases approximately 50\,nm (about 35$\%$) with the increase in $K_{z}$ from zero to 237\,kJ/m$^3$.

Several authors have addressed the problem of dynamic switching of vortex core polarity. Lee et al.  \cite{Lee:2008} proposed a universal criterion for the critical velocity; these authors have considered that the critical velocity  $v_{\rm crit}$ (and also $r_{\rm crit}$) depends only on the intrinsic parameter of the magnetic material, namely, the exchange stiffness $A$:  $v_{\rm crit}\cong 1.66\gamma\sqrt{A}$.
\begin{figure}[h]
\includegraphics[width=1\columnwidth]{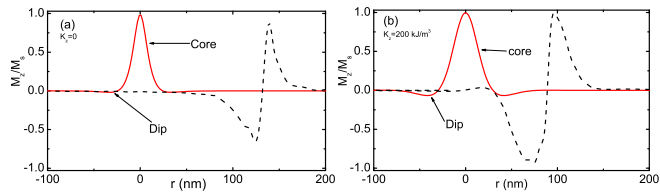}
\caption{Profile of the vortex core in equilibrium (continuous red line) and immediately before switching (dashed black line), showing the development of the dip, for a) $K_z=0$ and b) $K_z~=~200\,$kJ/m$^3$.}
\label{fig:Dip}
\end{figure}

Guslienko et al.\cite{Guslienko:2008} have explained the switching of the vortex core as due to the gyrofield that is proportional to the core velocity. This  field has a direction opposite to the core polarity, thus promoting the asymmetric increase of the dip. At the critical velocity, the dip magnetization becomes equal to $m_{z}=-1$,  and the vortex core switches, see Fig~\ref{fig:Dip}. Later Khvalkovskiy et al.\cite{Khvalkovskiy:2010} studied how the critical velocity is affected by a perpendicular magnetic field. These authors confirmed that the critical velocity is proportional to the radius of the vortex core, i.e., $v_{\rm crit}\approx \gamma_{0} R_{\rm c}$, as proposed earlier.\cite{Guslienko:2008}

\begin{figure}[h]
\includegraphics[width=0.9\columnwidth]{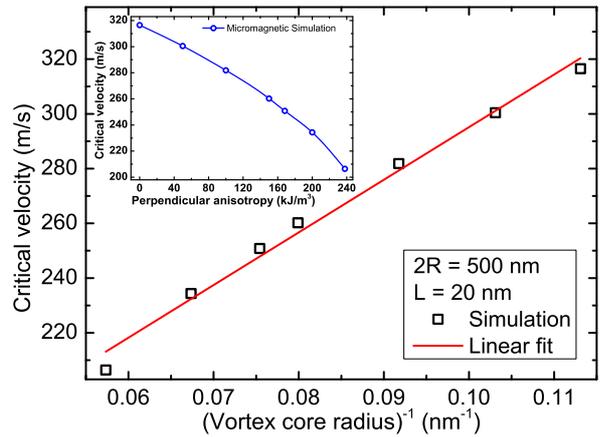}
\caption{Relation between the vortex core critical velocity ($v_{\rm crit}$) and the inverse of the core radius $R_{\rm c}^{-1}$ for a disk of $2R=500$\,nm and thickness $L=20$\,nm, varying the perpendicular anisotropy. The red continuous line is a linear fit of the form $v_{\rm crit}=A R_{\rm c}^{-1}$}.
\label{fig:VcriVsCore}
\end{figure}

Analogously, we varied the perpendicular anisotropy, in order to modify the core size, and finally study its effect on the critical velocity. These results are presented in Figure~\ref{fig:VcriVsCore}. The red continuous line represents the behavior of $v_{\rm crit}$ as a function of $R_{\rm c}$, showing that $v_{\rm crit}$ is roughly proportional to $1/R_{\rm c}$. In the inset of Fig.~\ref{fig:VcriVsCore}  the dependence of the critical velocity on $K_{z}$ is presented.


At a first glance, our results disagree with those found by Guslienko { et al.}\cite{Guslienko:2008} and Khvalkovskiy { et al.}\cite{Khvalkovskiy:2010}. However, it is worth noting that, while the perpendicular magnetic field acts differently on the core and on the dip, i.e., increasing one it will necessarily decrease the other one, on the other hand, the $K_z$ favors the increase of both core and dip, since the $K_z$ is uniaxial. This may be the cause of the opposite dependence on the critical velocity.

Therefore, we could conclude that the effect of the $K_z$ on the vortex core dynamics has a dominant role if compared to other aspects, such as the core size. We believe that the $K_z$ acts on the gyrofield, increasing it drastically and favoring the polarity switching. However, the details of the dependence of the vortex core dynamics on the $K_z$ are beyond the scope of the present work.


\section{Remote reversal of the core polarity}
\label{sec:twosdisk}
The dynamically induced magnetic interaction between spatially separated magnetic nanodisks has been studied in some previous works.\cite{Vogel:2010,Sukhostavets:2013,Jung:2010,Jung:2011} In the present work, we have used this same coupling to study the possibility of remote core polarity reversal in a pair of magnetic disks. In order to achieve this goal, our strategy was to excite, by applying a rotating magnetic field, just one of the disks. We chose the field amplitude, in such a way, to be just below the threshold of vortex core polarity reversal, ensuring a stationary core orbit with the largest possible radius, for a given anisotropy. We expected to induce the remote core reversal of the second disk (through dynamic magnetic coupling only)  in the case where the critical radius ($r_{crit}$) of this disk is smaller than $r_{crit}$ of the first one.
As we have shown above, for a single disk the $r_{crit}$ (as well as the critical velocity) depends inversely on the value of the perpendicular anisotropy, $r_{crit}$ presenting its largest value for $K_{z} = 0$. Therefore, in one disk of the pair (disk 1) the $K_z$ was always kept constant, and equal to zero ($K_{z} = 0$), in order to reach the maximum possible orbit. On the other hand we changed the $K_z$ (from $K_{z} = 0\  \rm{to} \  237 \,\rm{kJ/m}^3$) of the other disk (disk 2). It is important to emphasize that only one disk of the pair (disk 1) was excited with the rotating magnetic field, the other, (disk 2) being totally passive.

\begin{figure}[h]
\centering
\subfigure{\includegraphics[width=0.8\columnwidth]{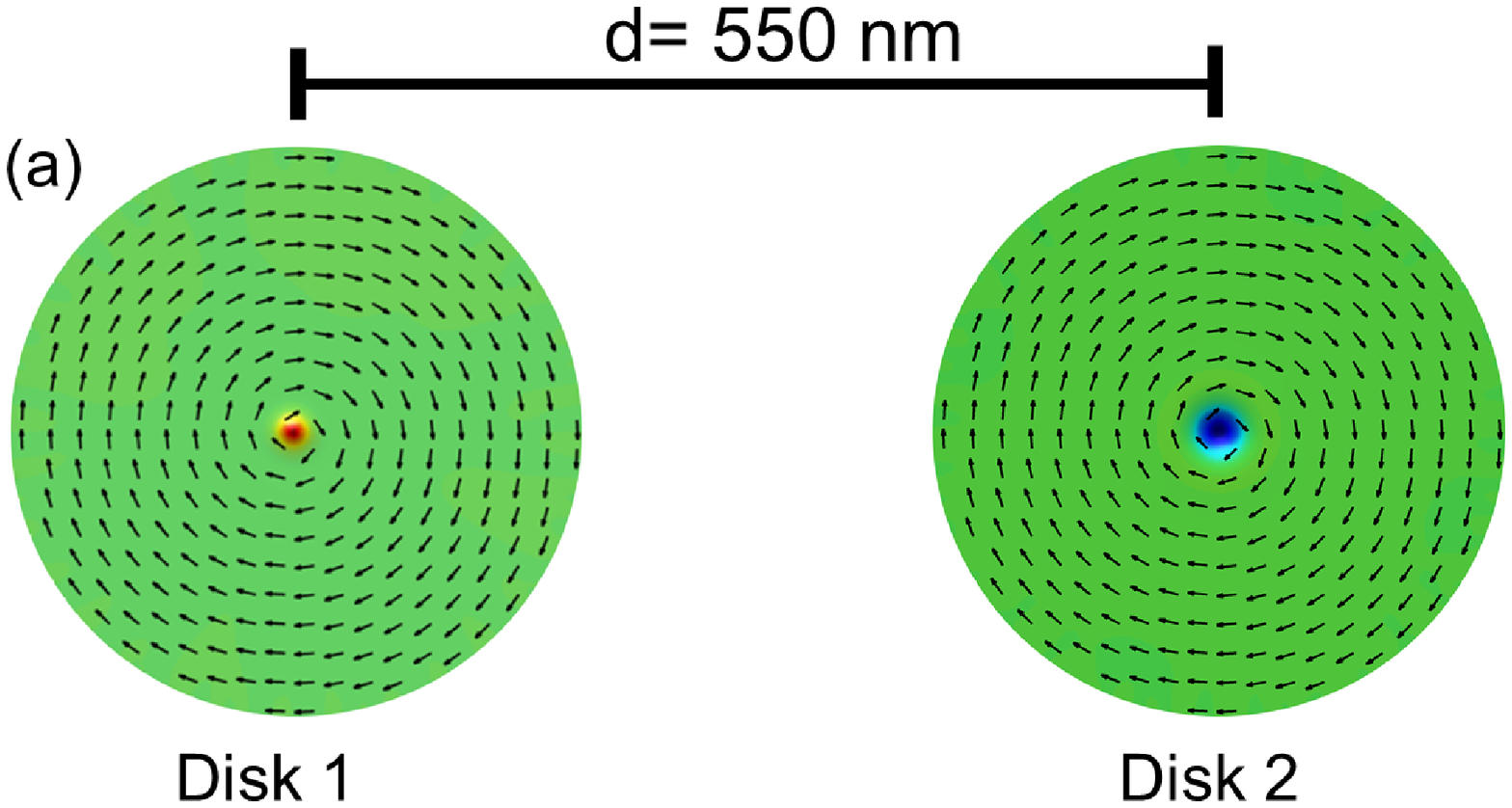} \label{Fig:DisksScheme}}
\subfigure{\includegraphics[width=0.9\columnwidth]{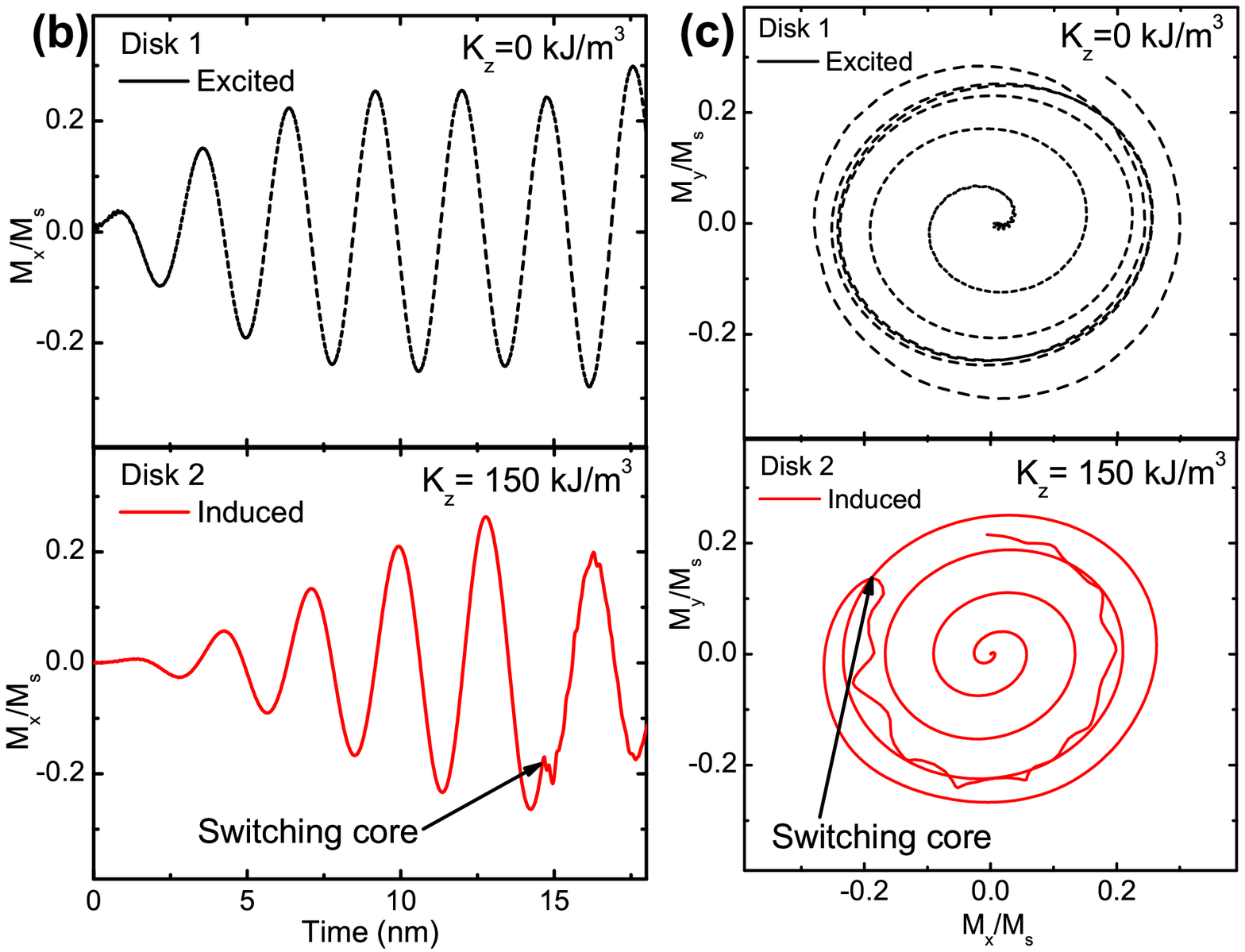}  \label{Fig:CorePosition}}
\caption{In {\bf  (a)} scheme a pair of disks with different perpendicular anisotropies. In {\bf (b)} we show that $M_{\rm x}$ of the excited disk 1 (black dashed line) has the same amplitude $M_{\rm x}$ as disk 2 (red solid line) induced by the motion of disk 1. In {\bf (c)} we show the gyrotropic motion for the excited disk (black dashed line) and the inversion of the motion of the core in disk 2 (red solid line), due to the inversion of its polarity.}
\label{Fig:CorePosition}
\end{figure}

In our simulations both disks have the same diameter, $2R=500$\,nm and the same thickness, $L=20$\,nm. The distance between the centers of the disks was maintained the same for all cases, $d = 550$\,nm (Fig~\ref{Fig:DisksScheme}). We observed that a pair of disks with the same polarity ($p_1p_2=1$) has a weaker coupling than the pair with opposite polarity ($p_1p_2=-1$); also, the circulation of the disks is irrelevant for the coupling. These findings are consistent with previous results.\cite{Garcia:2012,Sinnecker:2014}

As expected, in a pair of disks with $p_1p_2=-1$, separated by a distance $d=550$\,nm and with the same $K_z$, we observed that, by exciting only one disk, the amplitude of the oscillatory motion is approximately the same for both disks, with a phase difference. It was observed an evident energy transfer between the two disks. On the other hand, when we introduced the $K_z$ on disk 2, we could observe the switching of polarity of this disk, purely induced by the movement of disk 1, which creates a magnetic coupling between them, as we can see in Fig.~\ref{Fig:CorePosition}(b) and (c). The polarity reversal only takes place due to the fact that the critical radius of disk 2 is smaller than the radius for $K_{z}=0$, therefore, disk 2 is able to reach the critical condition, and then switch.


\begin{figure}[h]
\includegraphics[width=0.9\columnwidth]{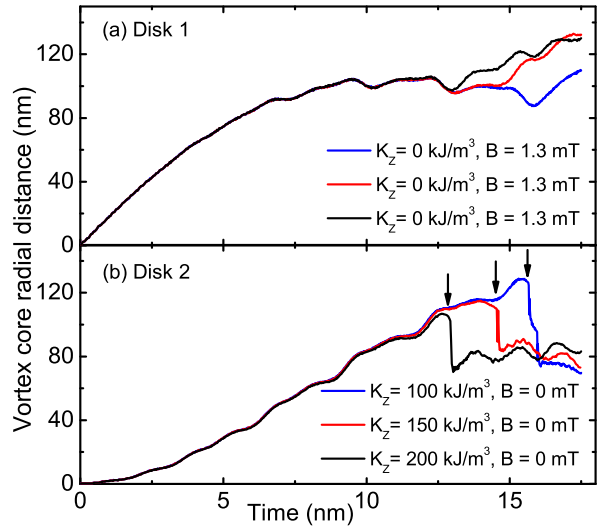}
\caption{Radial distance traveled by the vortex cores in the system of two disks. In {(a)} disk 1, with $K_{z}=0$, with applied rotating magnetic fields of  different amplitudes; {(b)} disk 2, with $K_{z}=200$\,kJ/m$^3$ no applied rotating field. The discontinuities in the curves, indicated by the arrows, show where there was a polarity change.}
\label{fig:discoduplo}
\end{figure}

Following this idea, we have been able to reverse indirectly the polarity of a vortex core, as it is shown in Fig.~\ref{fig:discoduplo}. Disk 1, with $K_{z}=0$ (in Fig.~\ref{fig:discoduplo}{(a)}) was excited with a given rotating field amplitude, with its natural frequency $\omega_0$, while for disk 2 the $K_z$ was varied. The discontinuities in the radial distance of the core of disk 2 characterize the moments when the core inverts. We see that the reversal occurs for disk 2, although the polarity of the excited disk is maintained. Disk 2 (see Fig.~\ref{fig:discoduplo}{(b)}) has not been excited by the rotating field, but the stray field from disk 1 induces the switching of its core polarity.

In Fig. \ref{fig:CoreRadialDistanceXkz} one could see that the radial distance of the vortex cores for a single disk before inversion shows the same time dependence, independently of the value of $K_z$. The same is true of the radial distance of the core in Disk 1 and Disk 2 in the interacting pair of disks, as seen in Fig. \ref{fig:discoduplo}. The shape of the curve, however, is modified in relation to the case of one single disk, as the motion of the vortex in Disk 1 affects the motion in Disk 2, and vice-versa.



The fact that one "master disk'' can be used to tune the polarity of the neighboring disks, without altering its own polarity, is relevant for future applications. This disk would have $K_{z}=0$ and the other would have different values for $K_{z}$, so that, if we employ $B=1.3$\,mT, for  ~14\,ns, for instance, only disks with $K_{z}=200$\,kJ/m$^3$ or larger would reverse their polarity. If B is applied for ~15\,ns, disks of $K_{z}$ up to 150\,kJ/m$^3$ would invert. We could therefore create an array of disks with different states of memory. Exciting disk 1 with a constant eigenfrequency $\omega_0$ and different elapsed times one could selectively switch neighbor disks with different anisotropies.

\section{Conclusions}
\label{sec:conclusions}

In summary, in this study we explored the influence of a perpendicular anisotropy on the vortex core polarity reversal in two systems: single disk and pair of interacting disks. In the case of a single disk we observed that the natural frequency of the vortex motion is changed with the presence of a $K_z$. We applied a rotating magnetic field with a frequency corresponding to the motion in the presence of $K_z$, and measured the critical reversal velocity. We showed that the critical velocity decreases with increasing $K_z$, or with increasing vortex core radius. The critical velocity is inversely proportional to the vortex core radius: $v_{\rm crit}\propto1/R_{\rm c}(K_{z})$ for disks with $K_{z} \neq 0$, therefore departing from a universal dependence given as $v_{crit}\cong 1.66\gamma\sqrt{A}$  in the literature \cite{Lee:2008}.


In a pair of disks with different $K_z$, their natural frequencies are also different. By exciting only the disk with $K_z=0$ on its natural frequency and varying the field amplitude, the gyrotropic motion of one vortex induces a similar motion on the other. For a pair of disks with different $K_z$, we can find a configuration such that we may reverse the polarity of the second disk before reversing the polarity of the disk being excited by the rotating magnetic field. This indirect, or induced, polarity reversal, may be used in magnetic vortex applications, such as vortex magnetic memories.

\begin{acknowledgments}
The authors would like to thank the Brazilian agencies CNPq, FAPERJ, FAPESP and CAPES.
\end{acknowledgments}

%
%
%
\end{document}